# GEDICorrect: A Scalable Python Tool for Orbit-, Beam-, and Footprint-Level GEDI Geolocation Correction

Leonel Corado[a, 1], Sérgio Godinho[a, b, 1], Carlos Alberto Silva[c], Juan Guerra-Hernández[d], Francesco Valério[a, e, f], Teresa Gonçalves[b, g], Pedro Salgueiro[b, g]

[a] MED – Mediterranean Institute for Agriculture, Environment, and Development, CHANGE – Global Change and Sustainability Institute, Institute for Advanced Studies and Research, Universidade de Évora, Pólo da Mitra, Ap. 94, 7006-554, Évora, Portugal

[b] EaRSLab – Earth Remote Sensing Laboratory, University of Évora, 7000-671 Évora, Portugal

[c] Forest Biometrics, Remote Sensing and Artificial Intelligence Laboratory (SilvaLab), School of Forest, Fisheries, and Geomatics Sciences, University of Florida, PO Box 110410, Gainesville, FL 32611, USA

[d] Forest Research Centre, Associate Laboratory TERRA, School of Agriculture (ISA), University of Lisbon, Lisboa, Portugal

[e] CIBIO, Centro de Investigação em Biodiversidade e Recursos Genéticos, InBIO Laboratório Associado, Campus de Vairão, Universidade do Porto, 4485-661 Vairão, Portugal

[f] BIOPOLIS Program in Genomics, Biodiversity and Land Planning, CIBIO, Campus de Vairão, 4485-661 Vairão, Portugal

[g] VISTA Lab, Centro Algoritmi \ LASI, Institute for Advanced Studies and Research, University of Évora, Colégio Luís Verney, 7000-671 Évora, Portugal

[1] These authors contributed equally to this work.

**Highlights:**

- Introduces *GEDICorrect*, a Python framework for GEDI geolocation correction.
- Implements orbit-, beam-, and footprint-level correction strategies.
- Enables testing of multiple waveform matching methods and metrics.
- Allows pre-selection of high-quality footprints for reliable corrections.
- Achieves ~2.4× faster single-process execution compared with the *GEDI Simulator* baseline.
- Scales efficiently with parallel processing, reducing total runtime to ~4,3 h on 24 cores (~19.5× overall speedup).

**Abstract:**

Accurate geolocation is essential for the reliable use of GEDI (Global Ecosystem Dynamics Investigation) LiDAR data in footprint-scale applications such as aboveground biomass modeling, data fusion, and ecosystem monitoring. However, residual geolocation errors arising from both systematic biases and random ISS-induced jitter can significantly affect the accuracy of derived vegetation and terrain metrics. The main goal of this study is to develop and evaluate a flexible, computationally efficient framework (*GEDICorrect*), that enables geolocation correction of GEDI data at the orbit, beam, and footprint levels. We present *GEDICorrect*, an open-source Python framework that enables geolocation correction using multiple methods, waveform matching, terrain matching, and relative height (RH) profile matching, implemented within a flexible, parallelized processing environment. The framework integrates existing *GEDI Simulator* modules (*gediRat* and *gediMetrics*) and extends their functionality with flexible correction logic, multiple similarity metrics, adaptive footprint clustering, and optimized I/O handling. We applied *GEDICorrect* to a heterogeneous Mediterranean woodland in southern Portugal to assess its performance across correction levels and computational configurations. Using the Kullback–Leibler divergence as the waveform similarity metric, *GEDICorrect* improved canopy height (RH95) accuracy from $R^2 = 0.61$ (uncorrected) to 0.74 with the orbit-level correction, and up to $R^2 = 0.78$ with the footprint-level correction, reducing RMSE from 2.62 m (rRMSE = 43.13%) to 2.12 m (rRMSE = 34.97%) at the orbit-level, and 2.01 m (rRMSE = 33.05%) at the footprint-level. Terrain elevation accuracy also improved, decreasing RMSE by 0.34 m relative to uncorrected data and by 0.37 m compared to the *GEDI Simulator* baseline. In terms of computational efficiency, *GEDICorrect* achieved a ~2.4× speedup over the *GEDI Simulator* in single-process mode (reducing runtime from ~84 h to ~35 h) and scaled efficiently to 24 cores, completing the same task in ~4.3 h: an overall ~19.5× improvement. *GEDICorrect* offers a robust and scalable solution for improving GEDI geolocation accuracy while maintaining full compatibility with standard GEDI data products. Its design enables researchers to test and compare alternative correction strategies and waveform similarity metrics, providing a flexible platform for refining spaceborne LiDAR geolocation methods and enhancing the precision of vegetation and terrain characterization worldwide.

# 1. Introduction

The ability to quantify vegetation structure is essential for understanding terrestrial ecosystems, assessing carbon stocks, and addressing global environmental challenges such as biodiversity loss and climate change (Dubayah et al., 2020). Among remote sensing technologies, Light Detection and Ranging (LiDAR) has emerged as a powerful tool for acquiring high-resolution, three-dimensional (3D) data on vegetation structure and terrain elevation (Guo et al., 2021; Lefsky et al., 2002; Valbuena et al., 2020). Its capability to penetrate forest canopies and capture detailed vertical vegetation profiles makes LiDAR invaluable for applications such as biomass estimation, habitat modeling, and fire risk assessment (Martin-Ducup et al., 2025; Moudrỳ et al., 2022; Silva et al., 2017).

In recent years, spaceborne LiDAR missions have expanded the reach of this technology, enabling near-global data collection at high spatial and temporal resolutions. In 2018, NASA launched two significant spaceborne LiDAR missions: the Ice, Cloud, and Land Elevation Satellite-2 (ICESat-2) (Neumann et al., 2019) and the Global Ecosystem Dynamics Investigation (GEDI) mission (Dubayah et al., 2020). Both missions have been continuously collecting and delivering extensive LiDAR datasets at a near-global scale, presenting an unprecedented opportunity to assess and estimate key vertical vegetation metrics and biomass across large areas, free of cost, and with high temporal frequency (Burns et al., 2024; de Conto et al., 2024; Dubayah et al., 2022; Hunter et al., 2025; Lang et al., 2023; Potapov et al., 2021; Saatchi and Favrichon, 2023).

GEDI, mounted on the International Space Station (ISS), is the first spaceborne LiDAR system specifically designed to globally measure and monitor the three-dimensional structure of vegetation and topography. It provides crucial insights into Earth's carbon storage, ecosystem structure, and biodiversity. GEDI collects high-resolution waveform data both day and night, continuously covering the Earth's land surfaces between 51.6° N and 51.6° S latitudes, encompassing the Earth's tropical and temperate forests. The sensor operates with three main lasers, generating eight parallel beams (four "coverage" beams and four "full power" beams) for surface observations. These beams illuminate an area on the Earth's surface equivalent to a circle of approximately 25 meters in diameter, known as the footprint (Dubayah et al., 2020).

However, as with all spaceborne LiDAR systems, GEDI's potential is often limited by the need to correct for geolocation errors, which arise from both systematic (i.e., consistent system-level biases) and non-systematic errors (i.e., random) sources (e.g., Xu et al., 2023). Systematic errors, typically assumed to remain constant within the same orbit (Mitsuhashi et al., 2024), stem from various factors, such as instrument calibration inaccuracies, spacecraft attitude and orbital uncertainties, GNSS positioning errors, laser pointing deviations, and atmospheric delays (e.g. Luthcke et al., 2002; Luthcke et al., 2019; Wang et al., 2018; Xu et al., 2023, Zhao et al., 2022). Among these, misalignments of the platform attitude and the laser pointing inaccuracies have been documented as two of the most significant contributors to geolocation error (e.g. Luthcke et al., 2000; Wang et al., 2018; Xu et al., 2024). In contrast, non-systematic errors, which are random components that differ from one laser shot to another, are primarily influenced by the platform operating environment and variable surface conditions (e.g. Mitsuhashi et al., 2024; Xu et al., 2023). The ISS hosts a wide array of mechanical systems, such as solar panels, motors, centrifuges, fans, pumps, and compressors, that generate structural vibrations over a wide range of frequencies (McPherson et al., 2015; Nelson, 1994; Su et al., 2024). Even routine activities by the ISS crew, such as exercising and moving between modules, induces mechanical vibrations throughout the platform (McPerson et al., 2015). These sources of vibrations make GEDI laser beams susceptible to deviations from their intended target, leading to pointing jitter and, consequently, additional geolocation inaccuracies (Hancock et al., 2019; Mkaouar et al., 2025; Su et al., 2024).

Recent studies highlight the persistence of these errors. Shannon et al., (2024) reported a systematic geolocation bias of ~9.6 meters in GEDI data, consistent with the ~10.3 meters geolocation error in Version 2 data (Beck et al., 2021). However, they also observed a substantial variation at the footprint-level, supporting the existence of random, non-systematic geolocation errors. This footprint-level variability is further addressed by Schleich et al., (2023), who demonstrated that small temporal clusters of footprints exhibit locally coherent shifts. By calculating the mean offset of each cluster, they were able to capture and reduce the impact of ISS vibrations on geolocation errors. Together, these studies highlight that, even with the improved systematic geolocation accuracy in GEDI Version 2 data, additional refinements are still possible. Indeed, one of the primary challenges in using spaceborne LiDAR data, particularly from GEDI, is its geolocation error, where the reported coordinates may not precisely correspond to the exact laser measurement location but rather to a nearby location in the surrounding area (Tang et al., 2023).

Efforts to improve GEDI's geolocation accuracy have been developed and implemented by the scientific community (e.g. Hancock et al., 2019; Mkaouar et al., 2025; Quirós et al., 2021; Schleich et al., 2023; Shannon et al., 2024; Xu et al., 2023; Xu et al., 2025). Notably, the GEDI Science Team developed the *GEDI Simulator* (Hancock et al., 2019), an open-source framework that includes a tool (*collocateWaves*), that uses small-footprint Airborne Laser Scanning (ALS) data to simulate GEDI waveforms and performs geolocation correction by aligning simulated and reported waveforms. The simulator also comprises two other key components: i) *gediRat*, which generates simulated waveforms from ALS data for specified coordinates; and ii) *gediMetrics*, which extracts vegetation and terrain metrics, including relative height (RH) profiles and ground elevation, from the simulated waveforms. The *collocateWaves* program applies an orbit-level correction strategy, assuming a systematic geolocation error across the entire orbit and calculating a single offset vector for all footprints within that orbit.

While the *GEDI Simulator* has been widely used, its reliance on orbit-level correction presents certain challenges. The assumption of uniform systematic errors across an orbit (Tang et al., 2023) is unlikely to hold true in regions with high topographic variability or heterogeneous vegetation (Milenković et al., 2017; Roy et al., 2021). In fact, Tang et al., (2023) conclude that assuming a constant systematic offset along the orbit is a major simplification and seldom holds in practice. This limitation has prompted the development of more fine-grained approaches, such as beam-level corrections (Tang et al., 2023; Yang et al., 2024), which estimate offsets for individual beam tracks, and footprint-level corrections (Mkaouar et al., 2025; Quirós et al., 2021; Yang et al., 2024; Schleich et al., 2023; Xu et al., 2025), which calculates offsets for individual footprint or small groups of footprints. While these methods offer greater precision, they also demand increased computational resources and more flexible tools for effective implementation.

In this work, we introduce *GEDICorrect*, an open-source Python framework designed to address these challenges by building upon and extending the capabilities of the *GEDI Simulator* and existing geolocation methods. *GEDICorrect* integrates *gediRat* and *gediMetrics* from the *GEDI Simulator* while introducing new geolocation correction methods and evaluation criteria. It supports orbit-level, beam-level, and footprint-level corrections, allowing users not only to tailor the correction process to their specific research needs but also to gain deeper insights into the systematic or random nature of GEDI geolocation errors. By enabling users to test orbit-, beam-, and footprint-level correction approaches within a single framework, *GEDICorrect* provides a unique opportunity to analyze and compare these methods comprehensively. By incorporating different waveform matching methods and metrics, as well as terrain alignment, *GEDICorrect* represents a significant advancement in space-borne LiDAR geolocation correction using ALS data. Additionally, the framework leverages parallel processing to efficiently handle large-scale GEDI and ALS datasets, making it a scalable and adaptable solution for geolocation correction across diverse landscapes.

The main objective of this paper is to describe the architecture and functionality of *GEDICorrect* and to evaluate its performance across orbit-, beam-, and footprint-level correction strategies, using waveform matching as the basis for geolocation adjustments under different parallelization settings.

## 2. Methods

### 2.1 Footprint clustering method

As outlined in the introduction, the *GEDICorrect* tool was developed to estimate geolocation offsets at three levels: orbit-level, beam-level, and footprint-level. Among these, footprint-level correction is a central component of our framework, building on the approach proposed by Schleich et al. (2023). Their method demonstrated that GEDI footprint geolocation errors, while appearing random, often show temporal coherence due to the mechanical vibration characteristics of the ISS. Building on this insight, we adopt a similar strategy, where the horizontal geolocation offset for each target footprint is derived by testing a range of candidate locations for a temporally local cluster of footprints, then applying the optimal offset determined for that cluster to correct the target footprint (Figure 1). The target footprint is the centroid of the cluster. However, a key distinction in our approach lies in the flexibility of the offset evaluation method. While Schleich et al. (2023) focused solely on terrain matching, using high-resolution DEMs to identify the best alignment between GEDI reported ground elevation and DEM ground elevation, our framework generalizes this process by allowing multiple scoring strategies. In addition to terrain matching, *GEDICorrect* also supports waveform matching, which compares GEDI reported waveform shapes with reference waveforms derived from airborne LiDAR. This flexibility addresses a known limitation of terrain matching in topographically flat areas, where multiple candidate positions may share identical ground elevations; the method struggles to identify a unique optimal offset. Waveform matching, by contrast, incorporates vegetation structure, making it more discriminative even in areas with uniform terrain. This enhancement allows our method to be applied in a wider range of landscapes and acquisition conditions.

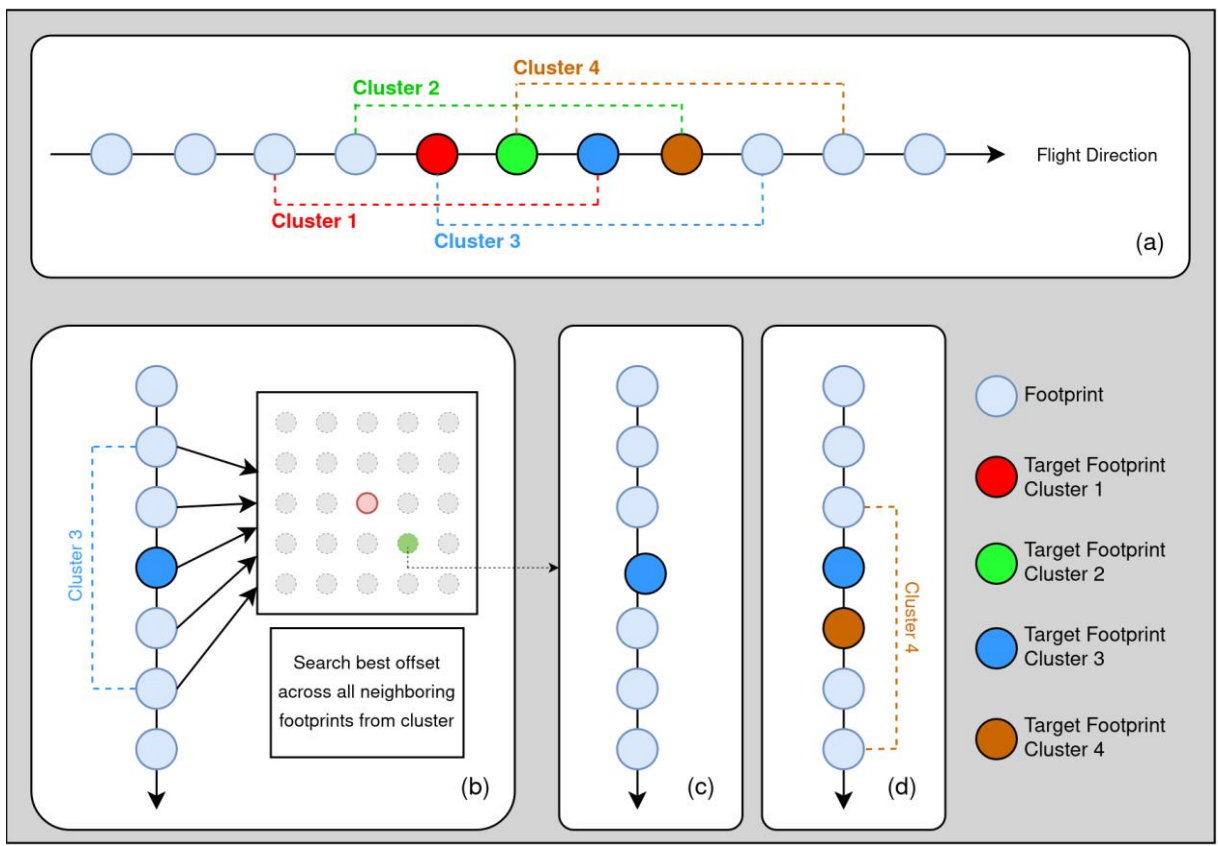

Figure 1 - Geolocation correction process at the footprint-level using Clusters: small groups of temporally correlated footprints.

Although GEDI emits laser pulses at a high frequency (242 Hz), structural vibrations of the ISS, particularly within the Japanese Experiment Module-Exposed Facility (JEM-EF) where GEDI is installed (Dubayah et al., 2020), occur at significantly lower frequencies, typically between 0.1 and 5 Hz (McPherson et al., 2015). This mismatch means that the platform's pointing deviation evolves slowly compared to GEDI's pulse rate, implying that groups of consecutive footprints, rather than individual ones, tend to share a common positional bias. As such, treating each footprint as having a fully independent error may overstate the randomness of the geolocation noise (Schleich et al., 2023). In this work, we refine the footprint-level correction strategy by explicitly modeling these short-term temporal correlations. We assume that GEDI footprints acquired over short time windows are subject to similar geolocation errors and can thus be clustered for local correction. This approach allows us to capture both the fine-scale variability and the temporally correlated pointing deviations introduced by ISS structural dynamics, offering a more physically realistic correction model than methods assuming either purely random or purely systematic error distributions.

To operationalize this, we group GEDI footprints into temporal clusters based on the known instrument's jitter frequency of ~5 Hz (Sipps and Magruder, 2023), which correspond to oscillation periods of ~0.2 seconds. In contrast, GEDI emits laser shots at 242 Hz, or roughly one footprint every 4.13 milliseconds. This mismatch in temporal scales implies that footprints acquired within short intervals (e.g., 0.2 seconds) are likely to share a relatively stable pointing error (Schleich et al., 2023). A 0.2-second cluster size would thus encompass ~48 consecutive footprints per beam (242Hz × 0.2s ≈ 48), equivalent to ~2.9 km along track. Schleich et al. (2023) demonstrated that such window size is short enough to avoid averaging across changes in ISS mechanical vibrations, while long enough to support robust local offset estimation. However, Sipps and Magruder (2023) noted that in regions with complex terrain and/or heterogeneous vegetation, shorter clusters can provide more reliable results, as they are more sensitive to instrument jitter and take advantage of local topographic or canopy structure variability to improve the match between reported and simulated waveforms. Based on these insights, and given that our case study area is structurally heterogeneous, we adopt a shorter cluster length of 8-12 footprints (~0.03 - 0.05 s). This choice balances temporal stability with the ability to leverage fine-scale landscape variability. The resulting cluster-level offset is then applied to target footprints (i.e. cluster footprint centroid), improving geolocation accuracy while preserving local spatial patterns.

### 2.2. Framework Design

The *GEDICorrect* framework is organized into three main units: i) Input; ii) Correction; and iii) Output (Figure 2). Its workflow follows a linear and user-friendly design, enabling researchers to adapt the geolocation correction process to their specific datasets and research objectives.

2.2.1. Input Unit

The Input Unit begins by performing an initial verification of the input files, which include merged GEDI L1B and L2A data products (see Section 4.2), along with a directory containing the intersecting ALS (*.las*) files. For each ALS file, the system processes the data to create a boundary polygon using one of two user-defined modes: i) Simple Bounding Box, which creates a rectangular boundary around the point cloud by determining the minimum and maximum (X, Y) coordinates from the ALS header; and ii) Convex Hull, which generates a convex hull surrounding the point data, creating a tight-fitting boundary (Andrew, 1979). While the latter offers greater precision and reduces errors during the simulation process, it is computationally more demanding. To optimize performance for subsequent runs, the Input Unit stores these boundaries in a Shapefile format during the first execution with the given ALS data. This pre-computation reduces the overhead of reading ALS data in future runs, streamlining the footprint correction process.

Once the ALS bounds are created, the framework loads the GEDI input files (in GeoPackage format) into *GeoDataFrames*. For each GEDI footprint, a square buffer is constructed around its centroid to identify intersections with the ALS data. The default size of this buffer is set to 50 meters, ensuring coverage of two whole footprints (which are 25 meters in diameter). Footprints whose buffers fall outside the ALS bounds are excluded to focus the correction process on areas where ALS point clouds exist. If any input files are corrupted or missing, it prompts the user to provide a new set of inputs and retry the footprint correction process. The result of the Input Unit is a list of GEDI footprints that are within ALS bounds.

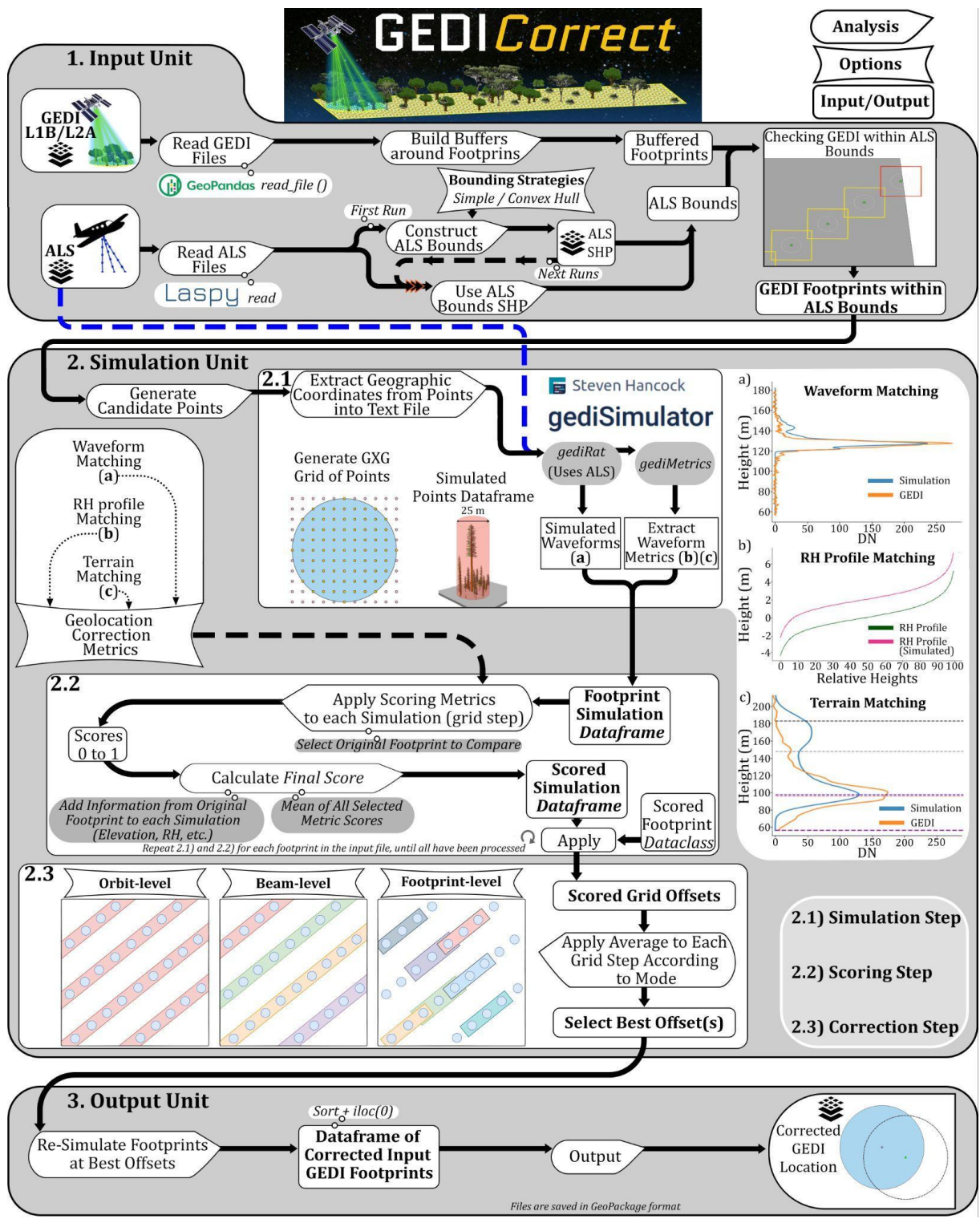

Figure 2 - Workflow of the *GEDICorrect* framework. The system operates through three main processing units: (1) Input Unit, where GEDI L1B/L2A and ALS data are read, buffered, and spatially intersected to identify footprints within ALS bounds; (2) Simulation Unit, where candidate offsets are generated and waveform simulations are performed using the *gediSimulator* module (Hancock et al. 2019). After simulation, users can select one of the three geolocation correction methods (waveform matching, RH-profile matching, or terrain matching) and choose among multiple similarity metrics for scoring the agreement between simulated and GEDI data. These scores are then used to estimate the best

horizontal offset at orbit-, beam-, or footprint-level; and (3) Output Unit, where corrected footprints are re-simulated and exported.

2.2.2. Correction Unit

After verifying each input GEDI file, every footprint undergoes a sequence of processing steps leading to its geolocation correction. The Correction Unit comprises three main steps: i) Simulation Step; ii) Scoring Step; and iii) Correction Step. This unit performs the horizontal geolocation correction based on user-defined parameters and is the computational core of *GEDICorrect*.

*Simulation Step*

Before the simulation step begins, the system generates a G x G grid of candidate offsets (in meters), spaced at regular intervals of size S (by default 1 meter), which is subsequently used during the simulation of candidate GEDI footprints. The simulation is accomplished through subprocess calls to the *gediRat* and *gediMetrics* programs within *GEDI Simulator* (Hancock, 2019). Based on the coordinates of the original GEDI footprint, the system uses the generated grid centered on the reported location and their respective geographic coordinates to store them in an ASCII file (Text file), which serves as input to the *gediRat* program. The *gediRat* program simulates GEDI waveforms for these coordinates using the provided ALS data, generating outputs in HDF5 format. Subsequently, the simulated waveforms (output of *gediRat*) are processed by *gediMetrics* to extract relevant RH metrics and other waveform properties, which are essential for subsequent scoring and selection of the best-corrected footprint position. The outputs from both programs are parsed and combined into a DataFrame, where each row corresponds to a simulated footprint. This *DataFrame* containing all simulated G x G candidate footprints around each original location constitutes the output of the Simulation Step and serves as the input to the Scoring Step. The provided ALS data for each *gediRat* subprocess call is restricted to the point cloud tiles that spatially intersect the 50 m buffer around the original footprint location (previously described in Section 2.2.1), typically resulting in the use of ~1 *.las* files per footprint. This targeted selection significantly reduces computational overhead by avoiding the need to load the full ALS dataset into memory, while still ensuring that all relevant terrain and canopy structure information is available for accurate waveform simulation.

To avoid potential impacts from land cover changes that may have occurred between the GEDI and ALS data acquisitions (e.g., tree cutting, wildfires), an additional filtering operation is performed during this step. This operation detects and discards footprints where the difference between the reported GEDI RH95 ($RH95_{orb}$), used here as a canopy height metric, and the mean RH95 from simulated GEDI values ($RH95_{sim}$) within the 30 x 30 m candidate grid exceeds a user-defined threshold. Such differences may indicate vegetation changes over time or inaccuracies in the simulation. The threshold is defined by the user based on the characteristics of the study area under analysis (the default value is 10 meters).

*Scoring Step*

The simulated footprints produced in the previous step are then evaluated in the Scoring Step, where a set of evaluation metrics, denoted as M, are applied to assess the similarity between the simulated and original footprints. Each *DataFrame* of simulated points, corresponding to a specific original footprint identified by its shot number variable, is compared to its original GEDI waveform from the "GEDI within ALS bounds" dataset (Section 2.2.1). Users can choose one or more of the available metrics (for more information about each metric, see Section 2.4), each generating a similarity score ranging from 0 to 1. If multiple metrics are selected, the final score for each simulated footprint is computed as the average of the individual metric scores. After calculating the final score, additional information from the original footprint is appended to each simulation, such as the original RH profile, waveform and terrain elevation. The output of the Scoring Step is an updated version of the simulation results, where each *DataFrame* now includes metric-specific scores, the aggregated final score, and relevant footprint metadata. During this step, we apply a dataclass named *ScoredFootprint*, where for each corrected footprint the minimal data is extracted to fulfill the requirements of the Correction Step, such as the shot number, beam, scores for each grid offset and the footprint *delta time* (used for clustering). By passing these compact objects instead of full *DataFrames*, we minimize inter-process communication overhead (described with more detail in Section 2.3.1). This structured output is then passed to the Correction Step, where the best-scoring offsets are selected for geolocation correction.

*Correction Step*

The scored results are then aggregated based on the correction strategy selected by the user: i) Orbit-level correction computes the mean score for each offset across the entire orbit and applies the offset with the highest mean score to all footprints; ii) Beam-level correction applies the same aggregation independently per beam, allowing different offsets for each beam; and iii) Footprint-level correction, which employs clustering algorithm described in Section 2.1, grouping temporally close footprints and selecting the offset with the highest aggregated score within each cluster. After determining the optimal offset for each footprint, the Correction Unit outputs a minimal data structure containing the shot number and its corresponding selected offset. This lightweight output is then passed to the Output Unit, where it is used to guide the resimulation process and generate the final corrected waveform data.

2.2.3. Output Unit

Once the optimal offsets are identified, the Output Unit performs a resimulation step at the selected offset positions to generate the final waveforms and relative height (RH) metrics for the corrected GEDI footprints. This resimulation follows the same procedures described previously in Section 2.2.2 (*Simulation Step* and *Scoring Step*). Before exporting the results, the simulated and corrected footprints are structured into a *GeoDataFrame*, which is then written to a *GeoPackage* file. This output file contains the waveform data, RH metrics, and associated geolocation information, ensuring compatibility with GIS software and enabling flexible downstream analysis. The framework allows users to process multiple GEDI orbit files, with each file undergoing the geolocation correction process independently. Once the process is completed for one orbit, the framework proceeds to the next file, repeating the process from the Correction Unit to the Output Unit, until all input GEDI files have been corrected.

## 2.3. Optimization Strategies

To enhance computational efficiency, scalability, and reduce the overall runtime, *GEDICorrect* incorporates several optimization strategies that accelerate the geolocation correction process without compromising the accuracy of simulation and scoring. These techniques include: i) Multiprocessing-based Parallelization, which enables simultaneous processing of multiple footprints and ensures efficient usage of computational resources (Figure 3); ii) Optimized memory management and I/O management during communication with the *gediRat* and *gediMetrics* programs, minimizing disk operations and reducing overhead; and iii) Selective reading and processing of ALS data, where only the relevant portions of point cloud files are accessed, significantly improving throughput, which was previously described in Section 2.2.2. (Simulation Step).

2.3.1. Parallel Processing

*GEDICorrect* employs a multiprocessing pool to process multiple GEDI footprints in parallel, substantially reducing the total computation time. Each footprint is handled as an independent processing unit, allowing it to pass through the entire Correction Unit (Simulation, Scoring, and Correction steps) concurrently with others footprints. This process is achieved by using a multiprocessing pool, which spawns N worker processes and assigns to each process a proportionally divided block of input footprints within the ALS data bounds for processing. To do this, the *pool.imap_unordered()* function from the *multiprocessing* Python library is used. To activate parallel processing, the user should enable the *'--parallel'* flag command and select the desired number of processes with the *'--n_processes'* command before executing *GEDICorrect*. If no number of processes is provided, the framework defaults to 8 processes. This simple configuration allows users to adapt processing workloads to their available computational resources.

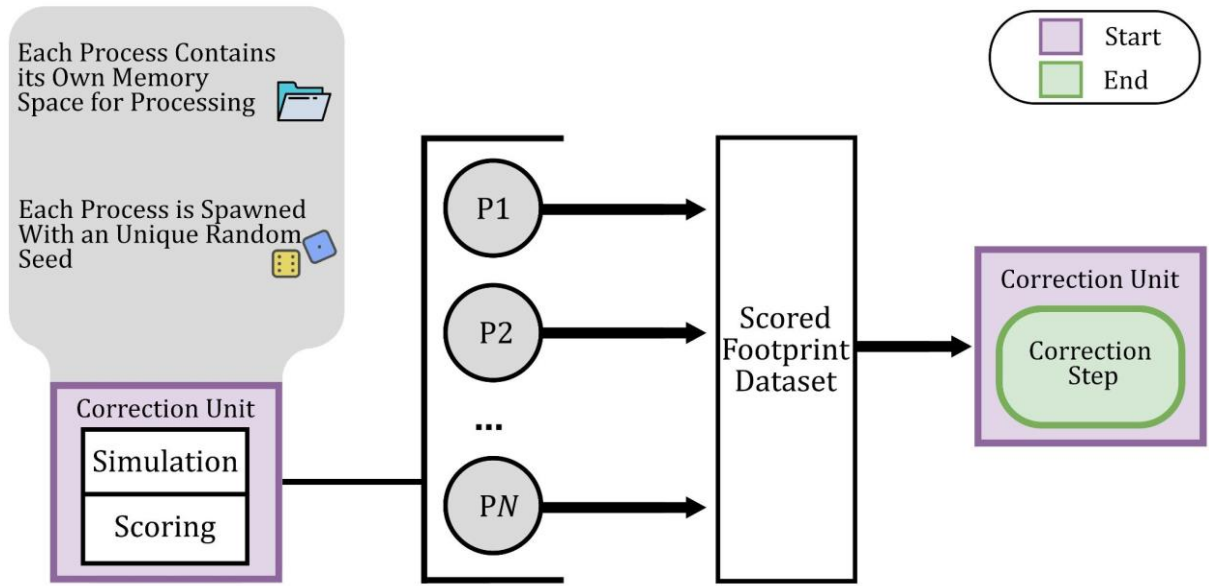

Figure 3 - Flowchart describing the Parallel Processing method of *GEDICorrect*. During Step 3, the scored footprints are transformed into minimal Data Structures (*ScoredFootprint* objects) before passing on the Correction Step, where all scored offsets are averaged according to the correction mode.

2.3.2. Directory and I/O Management

During the Simulation Step (Section 2.2.2), I/O handling across multiple calls to the *GEDI Simulator* programs (*gediRat* and *gediMetrics*) is managed via a *TemporaryDirectory*, where each process operates within its own isolated temporary directory to prevent program output conflicts and ensure that data generated by one process does not interfere with another one. For example, the geographic coordinates required by *gediRat* (described in Section 2.2.2) are saved in unique files within these directories. Moreover, output files generated by each process are prefixed with the process ID to further ensure isolation and avoid overwriting by other processes. Once the Correction Unit is complete, the temporary directories are deleted.

2.3.3. Selective ALS Data Access

A key performance optimization in *GEDICorrect* is its selective reading of ALS data. Instead of loading entire *.las* files into memory, which can contain millions of points and lead to a high computational overhead, the framework first reads only the header information to identify file extents and determine spatial intersections with GEDI footprints using the *laspy.read()* function. When the convex hull mode is selected (previously described in Section 2.2.1), the framework reads the necessary point data to compute the convex boundary enclosing the ALS footprint area. Although this operation is slightly more computationally demanding than the simple bounding box, it provides a more accurate spatial representation of the available ALS data, thereby reducing errors during waveform simulation. It then extracts only those ALS tiles that overlap a 50 m buffer around each footprint, typically requiring just one or two .las files per simulation. This targeted approach greatly reduces data-loading time and memory usage. Additionally, for subsequent runs on the same dataset, *GEDICorrect* can reuse precomputed boundary *shapefiles* generated during the initial Input Unit (Section 2.2.1), thereby bypassing the need to re-read point cloud headers and further reducing runtime. (This optimization was not enabled during the performance tests reported in this study.)

**2.4. Metrics for Geolocation Correction**

The *GEDI Simulator*'s *collocateWaves* program (Hancock, 2019) employs the Pearson correlation metric to compare original and simulated waveforms, assessing the linear relationship between their amplitudes at each bin, which was the method used by Blair and Hofton (1999). However, when correcting the geolocation of each footprint, the goal is to align the reported and simulated waveforms based on two critical features: i) the shape of the waveform curve and ii) the alignment or superposition of the curves. While Pearson correlation measures the strength of the linear relationship between two waveforms, it does not inherently account for their alignment, meaning that waveforms with similar shapes but shifted relative to one another can still yield high correlation values (e.g. Heersma et al., 2001; Rebonatto et al. 2017). This misalignment can prove undesirable in geolocation correction. Therefore, alternative approaches that explicitly consider both curve shape and alignment were included in *GEDICorrect*. Specifically, three matching methods were considered: i) Waveform matching; ii) Terrain Matching; and iii) RH Profile Matching, each containing its own set of criteria and metrics for calculating similarity scores (Figure 4).

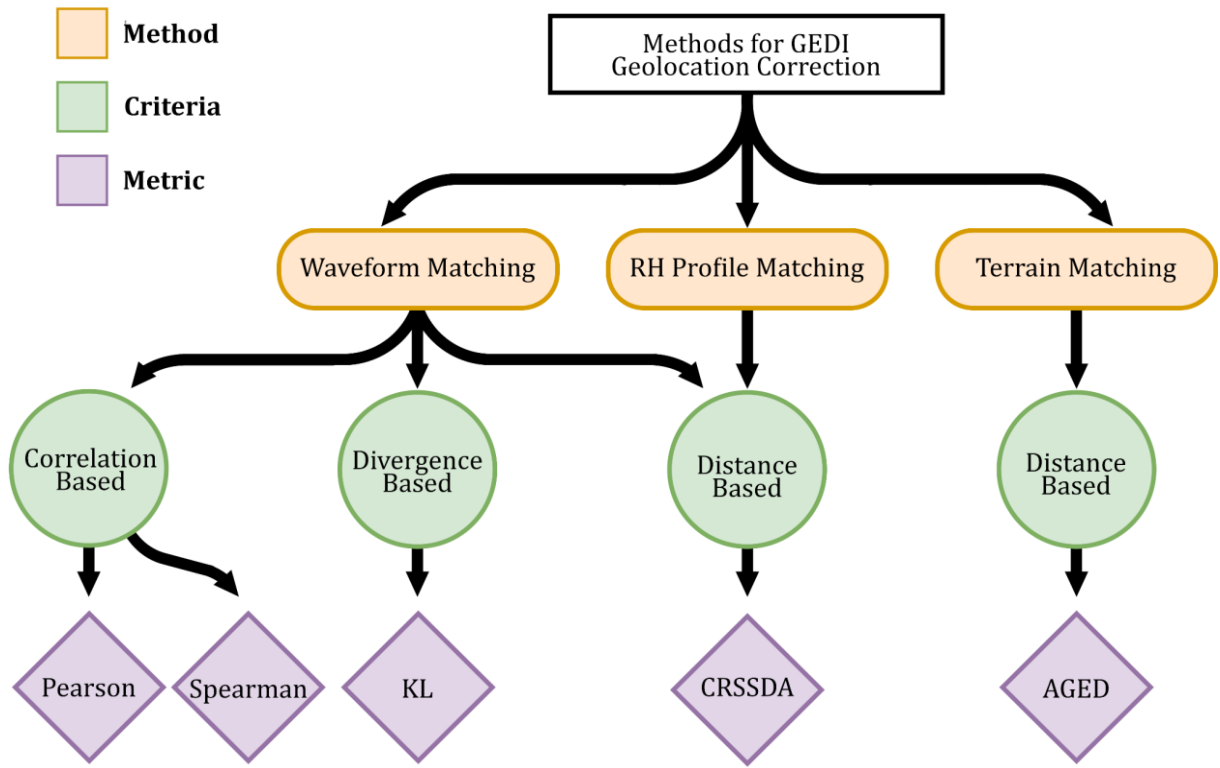

Figure 4 - Diagram illustrating the methods, criteria and metrics available for GEDI geolocation correction on *GEDICorrect* - Kullback-Leibler Divergence (KL); Pearson's and Spearman's Correlation; Curve Root Sum Squared Differential Area (CRSSDA) and Absolute Ground Elevation Distance (AGED).

Before executing the geolocation correction pipeline, the user selects the desired set of metrics (M) to calculate a similarity score. A list of metric names is available, including *'wave_pearson', 'wave_spearman', 'kl', 'wave_distance', 'terrain', and 'rh_distance'*. If multiple matching metrics are selected, the user can combine them by connecting each metric string with a *space (' ')* symbol.

2.4.1. Waveform Matching

For correlation-based methods, *GEDICorrect* implements both Pearson and Spearman correlation metrics, available as *'wave_pearson'* and *'wave_spearman'*, respectively. Pearson correlation measures the strength of the linear relationship between the amplitudes of the original and simulated waveforms across all bins. However, as previously mentioned, it is not sensitive to waveform shifts and does not account for alignment. In contrast, Spearman correlation assesses the rank-based relationship between the amplitudes, making it more robust to monotonic but non-linear relationships (Rebonatto et al. 2017). These correlation-based approaches are widely used due to their simplicity (Hancock et al. 2019; Blair & Hofton 1999) but may not fully capture both the alignment and shape similarity of waveforms. To overcome these limitations, *GEDICorrect* incorporates advanced curve similarity metrics inspired by the work of Zhou et al. (2016). Two key metrics were implemented: Curve Root

Sum Squared Differential Area (CRSSDA) and Kullback-Leibler (KL) Divergence, which are available in *GEDICorrect* as *'wave_distance'* and *'kl'* respectively.

CRSSDA (*'wave_distance'*, Equation 1) is an area-based measure that quantifies the similarity between a reported ($r_i$) and simulated ($s_i$) waveform by calculating the area between their curves. The method involves determining the squared difference between the two curves at each height bin (z), summing these differences across all bins, and taking the square root to compute the total differential area (Equation 1). A smaller CRSSDA value indicates a smaller difference between the reported and simulated waveform curves and, therefore, a higher similarity between the waveforms. On the other hand, KL Divergence (*'kl'*, Equation 2) (Kullback & Leibler, 1951) evaluates the dissimilarity between two probability distributions, making it a divergence-based metric. Unlike Pearson, which only measures linear dependence and is shift-invariant (Heersma et al., 2001), KL divergence is sensitive to subtle shifts in waveform distribution and shape (e.g. Fernandes et al., 2024). KL Divergence has been successfully applied in fields such as image pattern recognition, hyperspectral image classification, and waveform matching (Nayegandh et al., 2006; Olszewski, 2012; Zhou et al., 2016), making it a robust metric for GEDI waveform comparison. Since a waveform can be normalized as a probability distribution function, we used the KL divergence metric to assess the similarity between the reported ($r_i$) and the simulated ($s_i$) waveform using Equation 2. The KL metric measures the additional "information cost" required to represent the original waveform using the simulated waveform distribution. A smaller KL value indicates a closer match between the two distributions.

$$CRSSDA = \sqrt{\sum_{i=0}^{n} (r_i - s_i)^2} \qquad \text{Equation 1}$$

$$KL = \sum_{i=0}^{n} log(r_i / s_i) \times r_i \qquad \text{Equation 2}$$

2.4.2. Terrain Matching

Although GEDI provides highly accurate ground elevation estimates under most conditions, its performance is compromised in regions with steep slopes and highly heterogeneous topography (Adam et al. 2020; Fayad et al. 2021; Moudrý et al., 2024; Wang et al. 2022). Since each GEDI waveform captures the ground return, terrain matching serves as an effective starting point for aligning waveforms precisely. *GEDICorrect* implements a simple distance-based criteria, the Absolute Ground Elevation Distance (AGED), which matches the ground elevation from the original GEDI ($ZGr$, represented by the *elev_lowestmode* GEDI variable) to the ALS simulated ground elevation ($ZGs$) (*'terrain'*, Equation 3). The absolute value of the smallest elevation difference is granted a higher score. The rationale behind AGED lies in its ability to penalize larger deviations, which ensures that the most accurate matches contribute to the geolocation correction process.

$$AGED = |ZG_r - ZG_s| \qquad \text{Equation 3}$$

### 2.4.3. RH Profile Matching

The Relative Height (RH) profile is a key variable derived from GEDI waveforms, included in the GEDI L2A product (Dubayah et al., 2021a), that represents the cumulative distribution of laser energy reflected from different heights within a footprint. Each RH value corresponds to the height below which a specific percentage of the waveform energy is returned, ranging from the ground (RH0) to the top of the canopy (RH100) (Duncanson et al., 2022). This profile provides a detailed model of vegetation structure, canopy height, and internal heterogeneity. Since the RH profile can be reconstructed from both original and simulated waveforms (using *gediMetrics*, described in Section 2.2), aligning these profiles can serve as a robust method for geolocation correction. If the simulated RH profile closely matches the RH profile of the original footprint, it indicates a strong similarity in waveform alignment and, consequently, in the footprint's geolocation. For this, *GEDICorrect* employs an adapted CRSSDA metric for RH profile alignment ('*rh_distance*' , Equation 4). This metric evaluates the similarity between the original and simulated RH profiles by calculating the area between their respective curves across a range of RH intervals, from RH25 to RH100 in 5% increments. By assessing the cumulative alignment across these intervals, the metric captures the vertical structure of the vegetation within the footprint. A smaller 'rh_distance' value indicates a closer match between the reported ($rRH_i$) and simulated ($sRH_i$) RH profiles.

$$CRSSDA_{RH} = \sqrt{\sum_{i=0}^{n} (rRH_i - sRH_i)^2} \qquad \text{Equation 4}$$

### 3. Usage of GEDICorrect

To execute *GEDICorrect*, users must first install the framework following the instructions provided in the repository (*https://github.com/leonelluiscorado/GEDICorrect*). Once installed, the geolocation correction pipeline can be run using a single Python script (*gedi_correct.py*). This script initializes a *GEDICorrect* object and applies the selected correction method based on the user-defined settings. These settings, specified via command-line arguments, allow full customization of the correction process and are detailed in Table 1. For the orbit-, beam-, and footprint-level approaches, *GEDICorrect* evaluates candidate positions within a 30 x 30 meter grid, centered on the original footprint, by default. This grid consists of candidate points spaced at 1-meter intervals, ensuring fine spatial resolution. The 30-meter span in both the along-track and across-track directions ensures that simulations stay well within the average geolocation error of ~10 meters, as reported by Beck et al. 2021. Figure 5 illustrates several examples of how to execute *GEDICorrect* using different configurations. After running the program, the corrected footprints and associated output files are saved in the user-defined output directory.

Table 1 - Command options for *GEDICorrect* and their respective default values.

| Option | Description | Default Value |
| --- | --- | --- |
| *--granules_dir* | Specifies GEDI input file directory for batch correction. | ⌀ |
| *--input_file* | Specifies a single GEDI input file for correction. | ⌀ |
| *--las_dir* | Specifies the ALS files directory required for processing. Must overlap with input granule file(s). | ⌀ |
| *--out_dir* | Specifies the directory in which to save the output (either of 3 modes). | ⌀ |
| *--save_sim_points* | Flag option to save all simulated points around each footprint. | *False* |
| *--save_origin_location* | Flag option to save the original location simulated footprint. | *False* |
| *--mode* | Selects the footprint correction method between Orbit-level, Beam-level or Footprint-level, based on the list ["orbit", "beam", "footprint"]. | *"orbit"* |
| *--criteria* | Enumerates the set of criteria for best simulated footprint selection, based on the list ["wave_pearson", "wave_spearman", "kl", "wave_distance", "rh_distance", "terrain"]. | *"wave_pearson"* |
| *--grid_size* | Specifies the size of the grid for Orbit-level, Beam-level or Footprint-level correction methods. | *30 m* |
| *--grid_step* | Specifies the step size for the grid for Orbit-level, Beam-level or Footprint-level correction methods. | *1 m* |
| *--parallel* | Flag option to run *GEDICorrect* in parallel. | *False* |
| *--n_processes* | Specifies the number of processes to use for parallel processing. | *8* |
| *--time_window* | Specifies the time window in seconds to cluster footprints. Only usable in *footprint-level* correction *mode*. A *time_window* of 0 (zero) specifies that only clusters of size 1 will be used for correction. | *0.04* |
| *--als_crs* | Optional parameter, sets the EPSG code of the input ALS. | *None* |
| *--als_algorithm* | Set the ALS bounding algorithm. Defaults to 'convex', which builds a tight-fitting boundary. The other | *convex* |

| | | option, 'simple', creates a simple bounding box around the ALS. | |

```
(1)  python3  gedi_correct.py    --input_file "mygedifle.gpkg"
                                 --las_dir "./als_portugal"
                                 --out_dir "./correct_gedi/"
```
(1) Execute GEDICorrect with the default settings on a single GEDI file. Runs at the orbit-level using the Pearson's Correlation waveform matching

```
(2)  python3  gedi_correct.py    --input_file "mygedifle.gpkg"
                                 --las_dir "./als_portugal"
                                 --out_dir "./correct_gedi/"
                                 --mode "footprint"
                                 --time_window "0.125"
                                 --criteria "kl"
                                 --parallel
```
(2) Executes on a single GEDI file at the "Footprint-level" with a time window of 0.125 (~20 footprints) using the KL metric and parallel processing. Since no "n_processes" were assigned, it defaults to 8.

```
(3)  python3  gedi_correct.py    --granules_dir "./my_gedi_dir"
                                 --las_dir "./als_portugal"
                                 --out_dir "./correct_gedi/"
                                 --mode "beam"
                                 --criteria "wave_distance"
                                 --parallel
                                 --n_processes 6
```
(3) Executes on a directory of GEDI files to correct at the Beam-level using the CRSSDA metric on the waveforms. It also uses parallel processing with 6 processes

```
(4)  python3  gedi_correct.py    --granules_dir "./my_gedi_dir"
                                 --las_dir "./als_portugal"
                                 --out_dir "./correct_gedi/"
                                 --mode "footprint"
                                 --criteria "wave_spearman wave_distance kl"
                                 --parallel
                                 --n_processes 16
```
(4) Executes on a directory of GEDI files to correct at the Footprint-level using multiple metrics (Spearman Correlation, Waveform CRSSDA and KL). It also uses parallel processing with 16 processes.

Figure 5 - Demonstration of execution commands used to run *GEDICorrect*.

## 3.1 - Vertical Datum Difference between GEDI and ALS

GEDI data is referenced to the WGS84 ellipsoid, with elevation values provided relative to the EGM2008 geoid model (Dubayah et al. 2021b). In contrast, ALS data may be processed using a different vertical reference, such as a national geoid or an orthometric height system. These differences in vertical datums between GEDI and ALS datasets can introduce systematic elevation biases, potentially leading to inaccurate waveform simulations and geolocation correction results. To ensure consistency in the geolocation correction process, it is essential to align both datasets to a common vertical reference (Liu et al. 2021). To support this, *GEDICorrect* includes an auxiliary script that transforms GEDI elevation estimates to match the vertical reference used in the ALS data. This transformation is performed using a GeoTIFF file, where each pixel represents the elevation difference relative to the WGS84 ellipsoid. The use of spatially varying elevation differences ensures accurate elevation alignment across different locations. By applying this transformation, users can ensure that both datasets are referenced to the same vertical datum, thereby minimizing elevation discrepancies in subsequent analyses.

## 4. Experiments

To evaluate *GEDICorrect* and its geolocation correction strategies, we conducted experiments in a test study area in Portugal. The objective of this evaluation was not to provide an exhaustive comparison of all possible metrics, but rather to demonstrate the accuracy, flexibility, and usability of the *GEDICorrect* framework across different correction levels and computational settings. While *GEDICorrect* supports multiple similarity metrics, in this study we focused on KL divergence to illustrate the framework performance. For this, we assessed three geolocation correction approaches: i) orbit-level correction, in which a single coordinate offset is applied to all footprints within the orbit; ii) beam-level correction, where independent corrections are computed for each individual beam track; and iii) footprint-level correction, where each target footprint is corrected using the optimal offset derived from its local cluster (see Section 2.1 for details). For waveform matching, we applied the KL divergence metric to assess its effectiveness in aligning reported and simulated waveforms. To evaluate computational performance, we run the framework under different levels of parallelization (N = 1, 2, 4, 8, 16, 24) and measured execution time for each correction strategy. To assess geolocation accuracy, we compared the simulated RH95 values of the corrected footprints ($RH95_{sim}$) to the corresponding reported RH95 values ($RH95_{orb}$). However, RH95 alone may not fully capture waveform alignment, since different canopy structures can produce similar canopy top height values. To address this, we also considered the difference between RH95 and RH50 ($\Delta RH95 - 50 = RH95 - RH50$) as a complementary indicator. This metric reflects the vertical distribution of canopy returns and has been used as a proxy for the vertical distribution of vegetation biomass within the canopy (e.g., Garcia et al., 2010; Garcia et al., 2017; Jensen et al., 2008). The rationale is that the arrangement of vegetation material (e.g., branches and leaves) within the canopy directly influences the waveform amplitude, and therefore its overall shape and characteristics (e.g., Bruening et al., 2021; Hyde et al., 2005). By incorporating $\Delta RH_{95-50}$, we reduce the likelihood of accepting matches where canopy height is similar but the underlying waveform structure differs, thereby providing a more stringent test of geolocation correction performance (e.g. Oliveira et al., 2023). To evaluate geolocation accuracy, the R², RMSE, and rRMSE were calculated using the following equations:

$$R^2 = 1 - \frac{\sum_{i=1}^{n}(y_i - \hat{y}_i)^2}{\sum_{i=1}^{n}(y_i - \bar{y}_i)} \quad \quad \text{Equation 5}$$

$$RMSE\ (m) = \sqrt{\frac{\sum_{i=1}^{n}(y_i - \hat{y}_i)^2}{n}} \quad \quad \text{Equation 6}$$

$$rRMSE\ (\%) = \frac{RMSE}{\bar{y}} \times 100 \qquad \qquad \text{Equation 7}$$

In these equations, $y_i$ represents the reported GEDI values *i*, $\hat{y}_i$ is the simulated GEDI values *i*, and $\bar{y}_i$ is the mean of the reported GEDI values *i*.

In addition to RH-based accuracy assessment, we also conducted an independent evaluation of the *GEDICorrect* geolocation performance based on ground elevation differences. Specifically, we compared the GEDI reported ground elevation (*elev_lowestmode*) with the ALS-simulated ground elevation both before and after applying the geolocation correction at orbit-, beam-, and footprint-level correction strategies.

The experiments were conducted using a machine equipped with an Intel(R) Core(TM) i9-14900K processor, featuring 24 cores and 32 logical processors (threads), operating at a base clock frequency of 3.20 GHz. The system includes 128 GB of RAM and runs Windows 11 Home, with Ubuntu 24.04.1 LTS (GNU/Linux 5.15.167.4-microsoft-standard-WSL2 x86_64) provided through the Windows Subsystem for Linux (WSL). Additionally, the machine is equipped with an NVIDIA GeForce RTX 4070 SUPER GPU with 12 GB of GDDR6 memory.

## 4.1 Study Area and ALS Data

The study area extends approximately 25.2 km in length and 39.5 km in width in southern Portugal (lat. 38° 59' N, long. -8° 7' W), where a savannah-like evergreen oak woodland known as *montado* (in Portugal) or *dehesa* (in Spain) is the dominant ecosystem. These woodlands represent a traditional agroforestry system dominated by holm oak and/or cork oak and are characterized by a high spatial variability in tree density, typically with an understory mosaic of annual crops, grasslands, and shrublands (Joffre, et al., 1999; van Doorn and Pinto Correia, 2007). The ALS data was acquired by the Direção-Geral do Território (DGT) as part of the 2024 National LiDAR Campaign, and has a nominal laser pulse density of 10 points/m², using a Riegl VQ-780II-S sensor mounted on an aircraft. In total, 977 *.las* files (with a total of 604 GB in size), with the Coordinate Reference System (CRS) set to the EPSG code of 3763, were used for the geolocation correction.

## 4.2 GEDI Data

The framework requires merged GEDI L1B and L2A data products in *GeoPackage* (.gpkg) format, as both are essential for metric calculations within the Correction Unit. Specifically, L1B waveform data is used for waveform matching, while the L2A relative height (RH) profile is used for RH profile matching. To streamline the merging process, the framework includes a utility script (*align_l1b_l2a.py*) that automates the merging of L1B and L2A datasets. This script accepts input GEDI data in GeoPackage (.gpkg) format. GEDI orbits intersecting the study area between 2021 and 2023 were downloaded and processed using the GEDI-Pipeline (Corado and Godinho, 2024), an open-source tool that supports searching, downloading, and processing GEDI data from NASA's Land Processes Distributed Active Archive Center (LP DAAC) via NASA's Common Metadata Repository (CMR). The downloaded GEDI data, originally in HDF5 format, were spatially clipped to the study area and preserved in their native format for subsequent use with the *GEDI Simulator*. In parallel, the data were also converted to GeoPackage format for compatibility with *GEDICorrect*, with relevant variables extracted and stored in new files for subsequent analyses. For computational efficiency in simulation and scoring tasks for both frameworks, a subset of GEDI files was selected from the available orbits intersecting the study area. This subset includes 11 files, containing a total of 18,630 footprints. Given the inherent noise and uncertainty in GEDI measurements, a preprocessing step was implemented to ensure that only high-quality footprints were retained for geolocation correction. This filtering process relied on a set of quality metrics developed and recommended by the GEDI Science Team and the research community. The specific filtering criteria used in this study are summarized in Table 2. After filtering, a total of 8,316 high-quality footprints were selected for input into *GEDICorrect*.

Table 2 - Quality metrics and criteria used to select high-quality GEDI footprints.

| Criterion | Description |
|---|---|
| *degrade_flag* == 0 | Indicates a low probability of degraded geolocation under suboptimal operating conditions (Roy et al. 2021). |
| *quality_flag* == 1 | Indicates that the footprint meets quality criteria in terms of energy, sensitivity, amplitude, and real-time surface tracking (Hofton et al. 2019). |
| *solar_elevation* < 0 | This metric is utilized to determine whether GEDI footprint acquisitions occur during night or day. Only the nighttime acquisitions were retained for analysis, as indicated by a solar elevation angle less than 0 (Beck, et al., 2021). |
| *sensitivity* >= 0.9 | Sensitivity refers to the maximum canopy cover that the GEDI laser shots can penetrate, considering the Signal to Noise Ratio (SNR) of the waveform. Based on previous studies that assess the impact of sensitivity on GEDI footprint accuracy (V.C. Oliveira et al., 2023), in this work, only footprints with a sensitivity greater than 0.90 were selected. |
| *(RH95 >= 5 and num_detected_modes == 1)* | This custom filter ensures that in all footprints representing forests (RH95 higher than 5 meters), the waveform generated by GEDI measurements exhibits more than one mode. Typically, a tree's waveform contains at least two modes: one corresponding to the canopy and another to the ground. |
| *RH95* <= 30 | This custom filter aims to eliminate erroneous canopy height measurements resulting from various factors (such as electric lines, aerosols, etc.) that interact with the GEDI LiDAR signal. For this case in Portugal, trees above 30 meters are rare. |

| | elev_lowestmode - digital_elevation_model | <= 50 m | To eliminate footprints with erroneous ground detection, all footprints with an absolute difference between the elevation of the lowest mode (elev_lowestmode) and the TanDEM-X elevation at the GEDI footprint location (digital_elevation_model) greater than 50 meters were excluded from the analysis. |
|---|---|

## 5. Results & Discussion

This section presents the main results of *GEDICorrect*, organized into four parts: (i) canopy metrics, (ii) waveform-level matching, (iii) terrain elevation, and (iv) computational efficiency. Because *GEDICorrect* was developed as an extension of the *GEDI Simulator*, we include the Simulator as a reference baseline in our experiments. The aim is not to provide an exhaustive comparison between the two frameworks, but rather to demonstrate where *GEDICorrect* yields measurable improvements beyond the Simulator in terms of accuracy and computational efficiency. Our primary evaluation focuses on the difference between uncorrected GEDI data and *GEDICorrect*-corrected outputs, with the *GEDI Simulator* serving as a widely recognized benchmark for context.

5.1 Effects of geolocation correction on accuracy of canopy metrics

Figure 6 compares simulations at the reported GEDI footprint locations with those at the corrected locations using the *GEDI Simulator* framework. At the reported positions, the relationship between simulated and reported RH95 values resulted in an $R^2$ of 0.61 and an RMSE of 2.61 m. Applying the *GEDI Simulator*'s orbit-level correction improved performance, increasing $R^2$ to 0.72 and reducing RMSE to 2.22 m. This corresponds to a 0.39 m reduction in RMSE and a 0.11 increase in $R^2$, consistent with findings by Cárdenas-Martínez et al., (2025), who reported similar improvements when comparing GEDI on-orbit data with post-collocation results using Hancock's method in a Mediterranean study area in southern Spain. When using $\Delta RH_{95-50}$ as the performance indicator, we observed a similar pattern, with clear improvements in $R^2$ and reductions in RMSE after applying the geolocation correction. However, the absolute $R^2$ and RMSE values were lower than those reported for RH95. After applying the geolocation adjustment using the standard *GEDI Simulator* tool, the mean horizontal offset was 11.54 m, slightly above the nominal horizontal accuracy of ~10 m reported for GEDI Version 2 data (Beck et al., 2021).

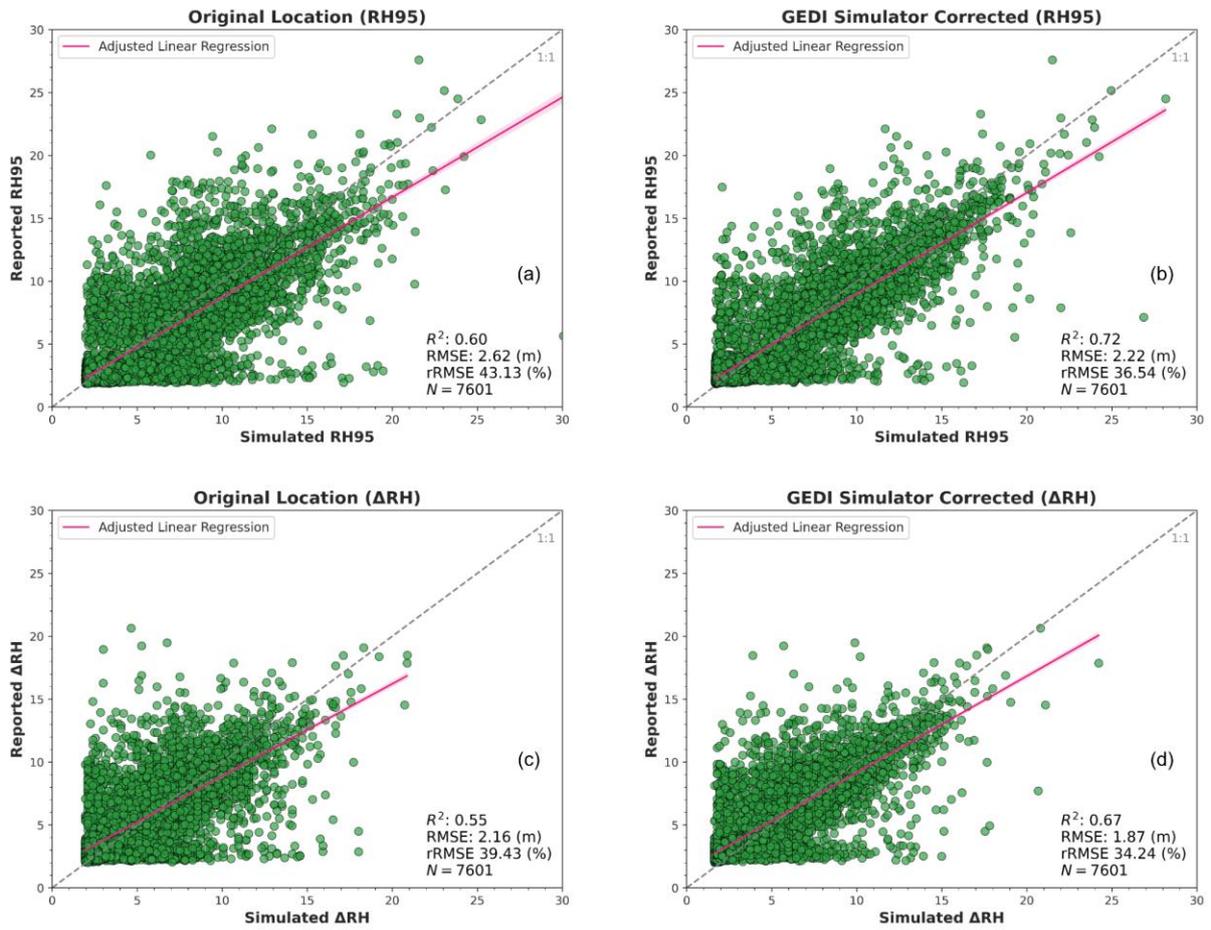

Figure 6 - Correlation plots between reported and simulated RH95 and $\Delta RH_{95-50}$ at original GEDI footprint locations and after applying *GEDI Simulator*'s orbit-level correction.

Figure 7 compares the three *GEDICorrect* correction strategies (orbit-, beam-, and footprint-level) using the KL divergence metric for waveform matching. For the cluster-based footprint-level correction, cluster sizes were defined using a time window of 0.04 s (~8-12 footprints per cluster). As expected, the footprint-level approach delivered the best results for this study area, achieving $R^2$ = 0.78 for RH95 and $R^2$ = 0.77 for $\Delta RH_{95-50}$, while orbit- and beam-level corrections produced slightly lower but comparable results (RH95: $R^2$ = 0.74; $\Delta RH_{95-50}$: $R^2$ = 0.70-0.71). When using RH95 as the performance indicator, all three approaches performed similarly, particularly for orbit- and beam level, suggesting that canopy top height alone is insufficient to reveal differences in correction quality. This limitation partly arises because RH95, while a robust proxy for canopy height, reduces the rich vertical information contained in a waveform to a single percentile value (e.g., Qu et al., 2018), which may obscure differences in canopy structure. In contrast, $\Delta RH_{95-50}$, which incorporates information on canopy vertical distribution (e.g., Jensen et al., 2008; Garcia et al., 2017), was more sensitive to waveform misalignments. By accounting for the differences between canopy top and median energy returns, $\Delta RH_{95-50}$ highlighted the added value of footprint-level corrections, particularly in heterogeneous savanna-like environments where canopy structure is highly variable (e.g. Carreiras et al., 2006; Li et al., 2023; Naidoo et al., 2012). Beyond $\Delta RH_{95-50}$, other vertical structure metrics such as Foliage

Height Diversity, available from the GEDI L2B product, have also been used to evaluate the impacts of geolocation errors (e.g., Cárdenas-Martínez et al., 2025) and represent a promising alternative for testing different correction strategies within the *GEDICorrect* framework.

In principle, footprint-level corrections should provide a clear advantage by capturing local variability in vegetation structure and terrain more precisely than broader orbit- or beam-level adjustments (e.g., Xu et al., 2025; Shannon et al., 2024). However, the improvements observed here were relatively modest, suggesting that external factors may overshadow their potential benefits. Specifically, residual geolocation errors induced by ISS vibrations and instrument jitter cannot be fully mitigated by correction algorithms (Nelson, 1994). Moreover, the similar performance across orbit-, beam-, and footprint-level corrections likely reflects the structural heterogeneity of the savanna-like ecosystem (*montado/dehesa*) that dominates the study area. In such landscapes, strong variability in tree density and structure at short spatial scales (Godinho et al., 2018; Rocchini et al., 2018; Dorado-Roda et al., 2021) reduce the sensitivity of waveform similarity metrics to geolocation offsets, as neighboring footprints may already capture distinct vegetation structures. This high spatial heterogeneity introduces complex vertical and horizontal vegetation patterns that are difficult to resolve with a cluster size of ~8 - 12 footprints (0.04 s time window, equivalent to clusters 480 - 720 m in length). Consequently, the choice of a 0.04 s clustering window, while relatively fine, may still be too coarse to capture the local variability in canopy and/or terrain conditions. This underscores the need to test smaller time windows and adaptive clustering strategies, which may help maximize the benefits of footprint-level corrections in heterogeneous environments (e.g. Sipps and Magruder, 2023). Complementary to this, Xu et al. (2025) proposed an approach that could further enhance clustering-based methods by applying footprint-level optimization after systematic corrections and clustering adjustments, focusing on reducing residual random errors through a small search area around each footprint to optimize waveform matching.

To summarize, when comparing the accuracy of canopy metrics between the original GEDI data (uncorrected) and the orbit-level correction using *GEDICorrect* framework, we found that *GEDICorrect* substantially improved canopy metrics accuracy, similar to, but slightly exceeding, the improvements obtained with the *GEDI Simulator*. For RH95, $R^2$ increased from 0.61 (uncorrected) to 0.74 with *GEDICorrect* (Figure 6a and Figure 7a), representing an improvement of 0.13. Similarly, RMSE decreased from 2.61 m (uncorrected) to 2.12 m with *GEDICorrect*. For $\Delta RH_{95-50}$, *GEDICorrect* also demonstrated a clear enhancement ($R^2$= 0.70, RMSE = 1.72) relative to the uncorrected data ($R^2$= 0.55, RMSE = 2.13 m). The mean horizontal offset derived from *GEDICorrect* at the orbit level was 10.79 m, closer to the nominal horizontal accuracy of ~10 m reported for GEDI Version 2 data (Beck et al., 2021) than the value obtained with the standard *GEDI Simulator*. These results confirm that *GEDICorrect* consistently improves canopy metrics accuracy and performs on par with, or slightly better than, its baseline *GEDI Simulator* framework, demonstrating its feasibility and usefulness for GEDI geolocation correction. The performance advantage observed for *GEDICorrect* likely stems from differences in data handling (input data) and offset estimation between the two frameworks. *GEDI Simulator* performs orbit-level corrections at L1B GEDI product using all available footprints within a given orbit, filtering only those with sensitivity < 0.9. The offset is then estimated by maximizing the Pearson correlation between reported and simulated waveforms across the candidate positions (Hancock et al., 2019). However, residual noisy or unreliable footprints may remain after this single filtering step, potentially distorting the correlation optimization and resulting in suboptimal offsets. In contrast, *GEDICorrect* offers greater flexibility by allowing users to pre-select high-quality footprints using a more comprehensive set of quality filters (Beck et al., 2021) before performing the correction (see Table 2). By relying on these higher-quality inputs, the framework minimizes the influence of poor-quality data and thereby enhances overall correction accuracy. Although in the present work *GEDICorrect* employs KL divergence rather than Pearson correlation, we believe that main source of its improved performance lies in the stricter waveform quality filtering, with the choice of similarity metric likely playing a secondary role. Nonetheless, disentangling the relative effects of waveform quality filtering and similarity metric selection deserves further investigation to more precisely quantify their respective contributions.

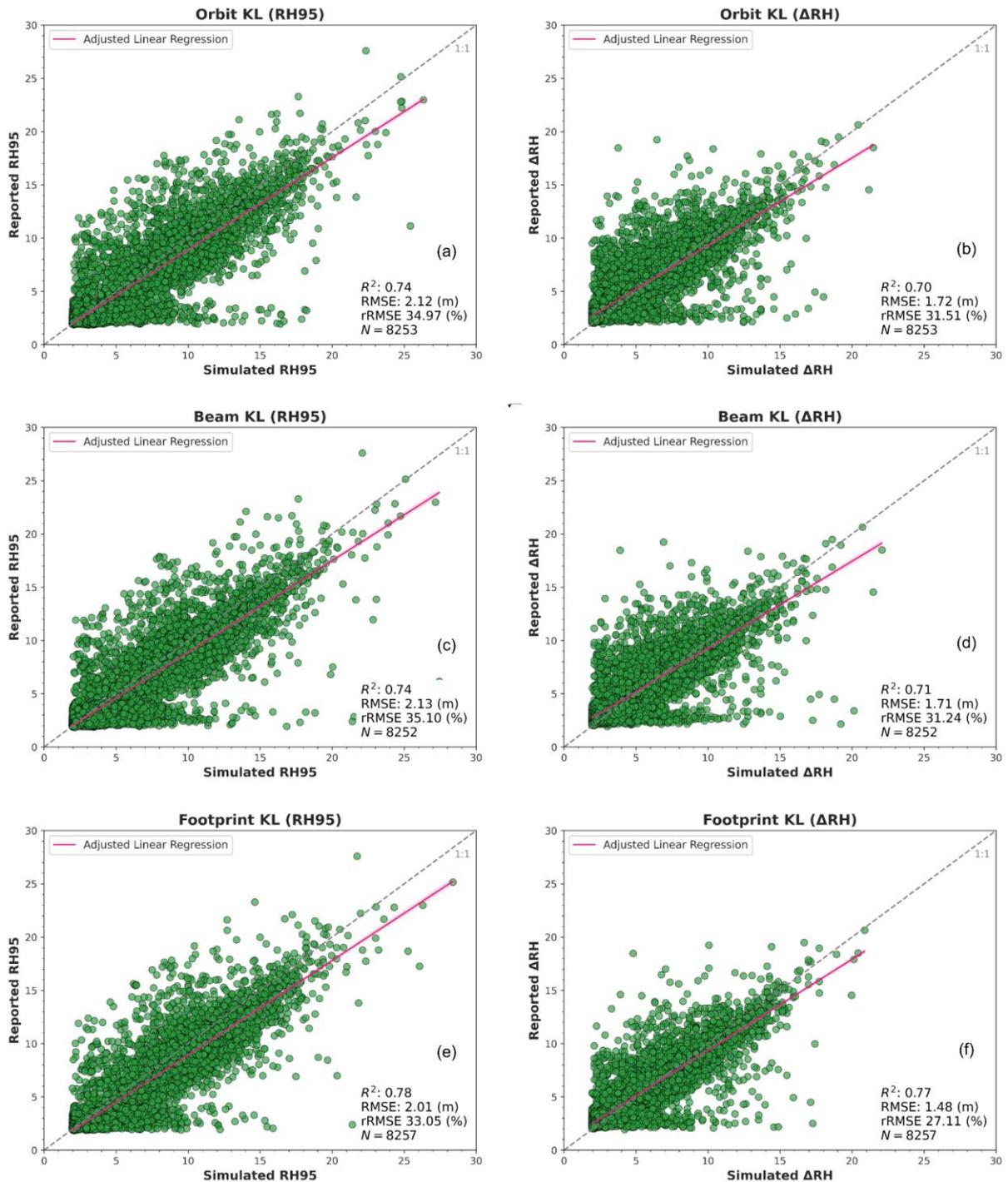

Figure 7 - Correlation plots between reported and simulated RH95 and $\Delta RH_{95-50}$ across orbit-, beam-, and footprint-level correction strategies using Kullback-Leibler Divergence (KL) as a waveform matching criteria.

## 5.2 Assessment of waveform-level quality matching

To complement the statistical results presented above, we examined individual waveforms to better understand the behavior of each correction approach. Figure 8 illustrates an example footprint (shot number '170811100200168667') in which reported and simulated waveforms are compared under different correction strategies.

Although the orbit-, beam-, and footprint-level approaches produced relatively similar $R^2$ values at the aggregate scale, the waveform comparisons revealed clear differences in alignment, particularly near the canopy top and ground returns. This contrast was most evident at the footprint-level (Figure 8d), where the simulated waveform followed the reported GEDI waveform more closely (Figure 8a). The usefulness of KL divergence for waveform matching has also been demonstrated by Zhou et al. (2016), who compared six similarity metrics using ICESat waveforms across different land-cover classes and found KL to deliver the highest average classification accuracy. This broader evidence supports our findings and underscores the value of KL divergence in discriminating subtle structural differences in waveform alignment. Therefore, careful inspection of individual waveforms remains essential for assessing the reliability of geolocation corrections. *GEDICorrect* addresses this need by implementing footprint-level waveform visualization, enabling users to validate corrections statistically and visually.

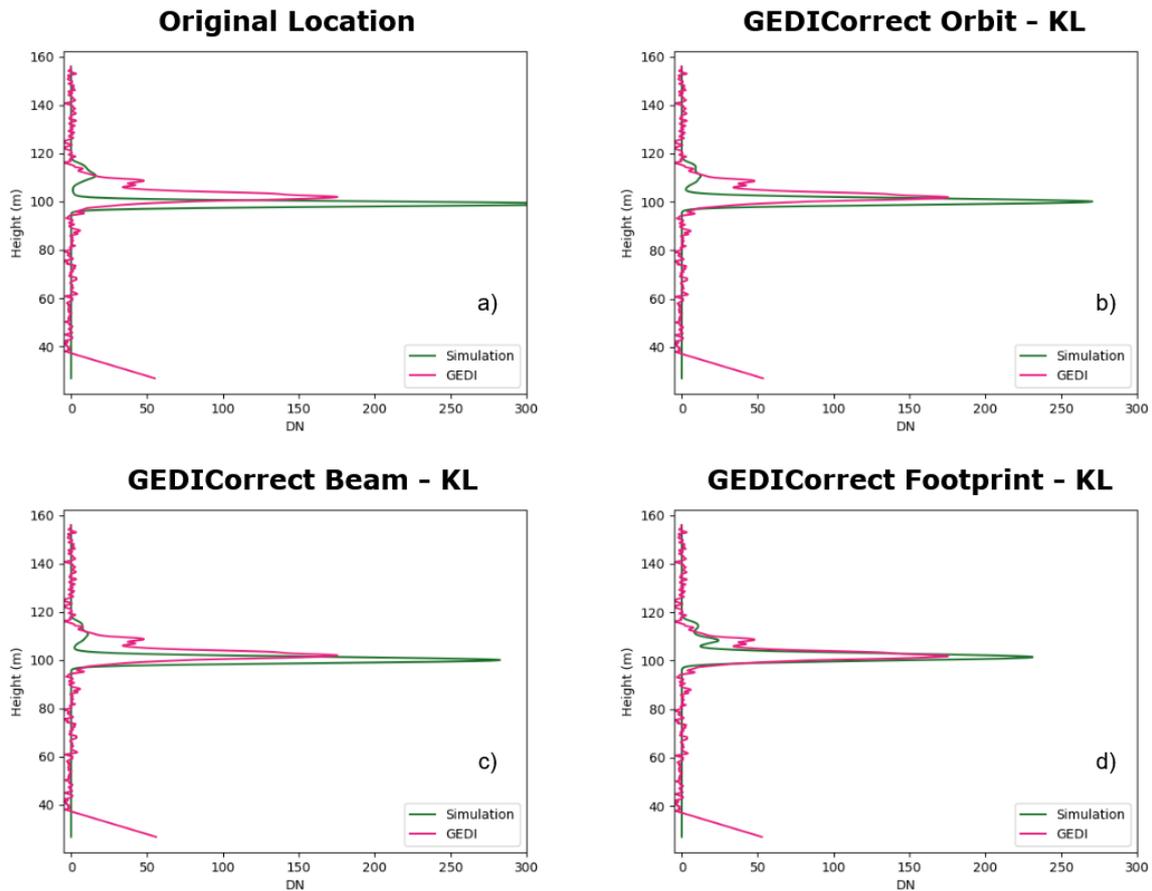

Figure 8 - Examples of waveform matching results across different correction strategies. Each panel shows the reported (pink) and simulated (green) GEDI waveforms for the same footprint.

5.3 Effects of geolocation correction on the accuracy of terrain elevation

To evaluate geolocation correction performance using terrain elevation as an indicator, we compared three cases: i) reported terrain height at the original GEDI footprint locations; ii) terrain height after *GEDI Simulator* correction; and iii) terrain height after applying *GEDICorrect* with the KL divergence metric at the footprint-level. In the *GEDI Simulator* case, RMSE increased slightly compared to the uncorrected footprints (from 1.83 m to 1.86 m; Figure 9a-b). This difference arises from the way *GEDI Simulator* applies vertical datum corrections. Specifically, *GEDI Simulator* computes a mean center-of-gravity (*CofG*) offset for the entire orbit and applies this value uniformly across all footprints (Hancock, 2019). While this uniform correction reduces broad systematic bias, it may fail to capture local terrain variations, leading to residual errors in areas of steep relief or uneven ground. This pattern is consistent with several studies that have examined the effects of terrain slope on GEDI performance (Adam et al. 2020; Fayad et al. 2021; Guerra-Hernández & Pascual, 2023; Wang et al. 2022). For example, Adam et al. (2020) showed that GEDI terrain and canopy estimates are more error-prone in heterogeneous or sloped environments, which likely explains the slightly higher RMSE observed in our study.

In contrast, *GEDICorrect* consistently produced the most accurate terrain results. After applying the KL-based footprint-level correction, terrain height RMSE decreased by 0.34 m relative to the original footprint positions and by 0.37 m compared to the *GEDI Simulator* framework (Figure 9c). This improvement in terrain elevation accuracy at the footprint level is consistent with Yang et al. (2024), who also compared orbit-, beam-, and footprint-level corrections and found that terrain estimates benefited most strongly from geolocation adjustment at the footprint scale. As described in Section 3.1, *GEDICorrect* leverages a geoid raster by applying the local geoid undulation value at each footprint. This localized adjustment enables closer alignment with ALS terrain data and reduces inconsistencies, particularly under complex terrain conditions. Taken together with the canopy results, these findings confirm that *GEDICorrect* improves both vegetation and terrain metrics consistently, underscoring its broader value as a geolocation correction framework.

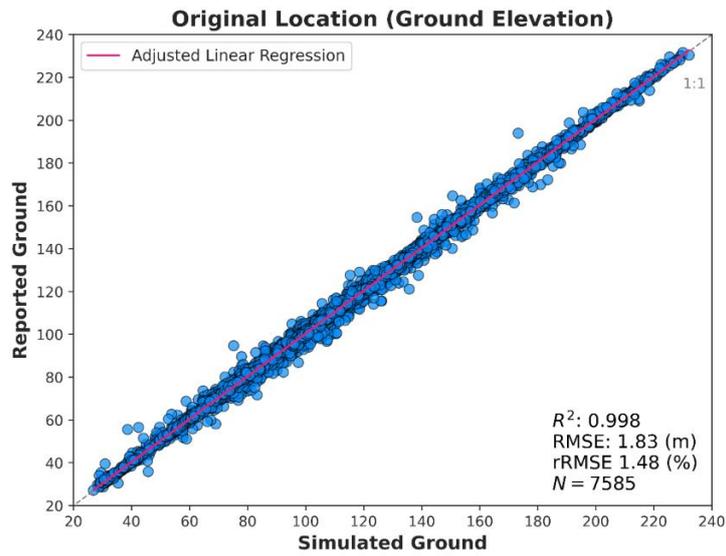
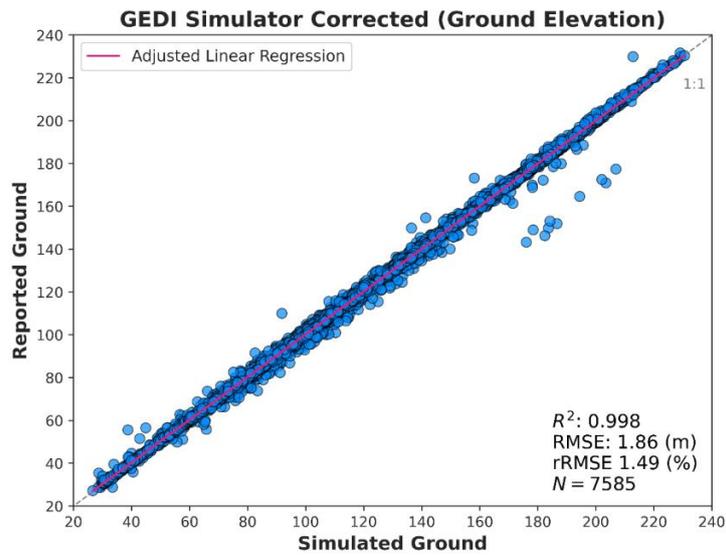
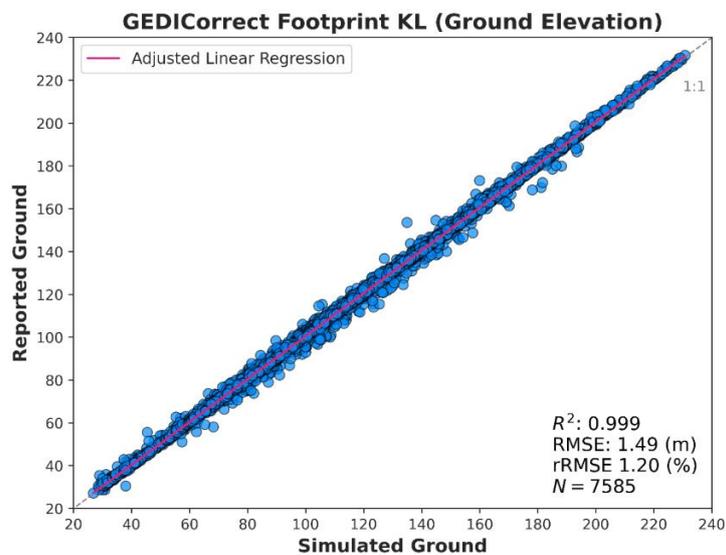

Figure 9 - Correlation plots between reported and simulated terrain elevation at the original location (a), after geolocation correction using *GEDI Simulator* (b) and *GEDICorrect* (c).

### 5.4. Computational performance evaluation

A key feature of the *GEDICorrect* framework is its support for parallel processing. In addition to testing different geolocation correction strategies (previously described in Sections 5.1-5.3), we evaluated the computational efficiency of *GEDICorrect* relative to *GEDI Simulator*. Table 3 reports the total processing time for both frameworks when conducting orbit-level correction. *GEDI Simulator* required approximately 84 hours to complete using single-process execution, whereas *GEDICorrect* completed the task in about 35 hours under the same setting, highlighting a substantial improvement in computational efficiency. This improvement stems mainly from *GEDICorrect*'s optimization strategies described in Sections 2.3.2 and 2.3.3.

Table 3 - Real time elapsed using single-process (P) execution on *GEDI Simulator* and *GEDICorrect* at orbit-level.

| Framework | P | Real Time Elapsed |
|---|---|---|
| *GEDI Simulator* | 1 | ~ 84 h |
| *GEDICorrect* | 1 | ~ 35 h |

For *GEDICorrect*, performance was further evaluated under different levels of parallelization. Figure 10 shows the elapsed time required to execute the *gedi_correct.py* script. For this study area and computational setup, processing time continued to decrease slightly as the number of parallel processes increased beyond 16, reaching its minimum at 24 processes (from ~279 minutes to ~256 minutes). This trend suggests that the framework scales efficiently with additional processes, although the performance gains could become progressively smaller at higher levels of parallelization due to overhead management between processes (Adefemi, 2024; Bhattacharjee et al., 2011; Roth et al., 2012). Computation times were comparable across the three correction strategies (orbit-, beam-, and footprint-level), indicating that the additional clustering step in the footprint-level approach did not impose a significant computational cost within *GEDICorrect*.

Although this study did not explicitly examine the influence of the number of simulated positions per footprint, this parameter is expected to influence both runtime and correction accuracy. Future work should therefore explore the trade-off between computational efficiency and geolocation accuracy by varying grid length and step size in waveform simulations.

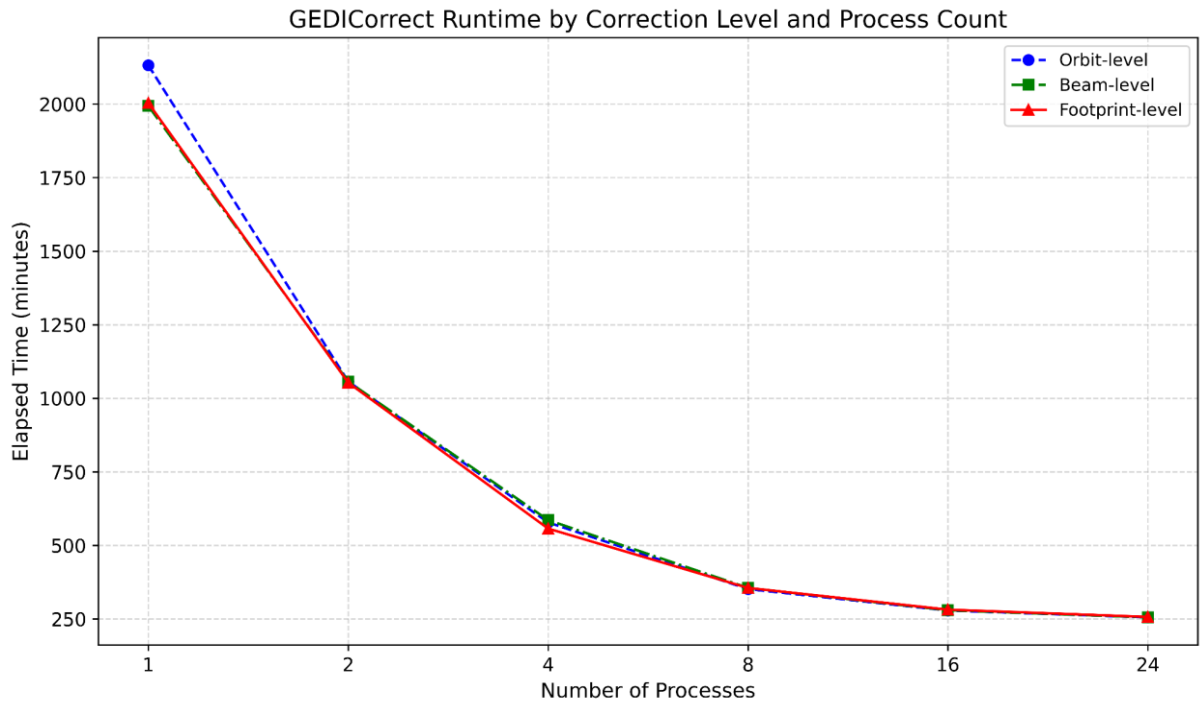

Figure 10 - Performance of *GEDICorrect* across different geolocation correction strategies (orbit-, beam, and footprint-level) as a function of the number of parallel processes used. As mentioned in Section 2.2.1, the ALS bounding algorithm was performed only once when using 1 process (orbit-level), whilst for other number of processes, the ALS bounding shapefile was used.

## 6. Conclusion

Geolocation uncertainty remains a major barrier to fully exploiting GEDI data for footprint-scale applications, including linkage to field plots for AGB modeling and data fusion with other remote sensing datasets. In this study, we introduced *GEDICorrect*, a Python framework that implements orbit-, beam-, and footprint-level correction strategies through waveform-based geolocation adjustment and parallelized processing. Our evaluation shows that *GEDICorrect* improves both canopy and terrain elevation accuracy relative to uncorrected footprints. Cluster-based footprint-level correction yielded the highest accuracy, particularly when assessed with structure-sensitive indicators such as $\Delta RH_{95-50}$. For terrain estimation, *GEDICorrect* consistently reduced RMSE compared to both uncorrected GEDI data and the baseline results produced by the *GEDI Simulator*, demonstrating its compatibility with the standard correction approach while improving overall efficiency. Importantly, *GEDICorrect* achieved these gains through a substantial increase in computational efficiency, running ~2.4× faster than the *GEDI Simulator* in single-process mode (~84 h → ~35 h) and scaling efficiently across 24 cores to complete in ~4,3 h on 24 cores, an overall ~19.5× improvement in processing time. Together, these results indicate that *GEDICorrect* fulfills its dual objectives: providing a flexible platform for testing geolocation correction strategies (orbit-, beam-, and footprint-level) and delivering computationally efficient tools suitable for large-scale analyses.

Looking forward, several improvements are possible. One priority is the implementation of adaptive clustering strategies, which would allow cluster size to vary depending on local terrain and land-cover complexity. As recent studies have shown (e.g., Schleich et al., 2023; Sipps & Magruder, 2023), the optimal cluster length depends on ISS jitter dynamics and landscape structure. Shorter clusters may better capture fine-scale offsets in complex terrain, whereas longer clusters can provide more stability in flatter or homogeneous regions. In addition, integrating pattern recognition techniques for waveform matching could further enhance correction accuracy by reducing the influence of random errors at the footprint level. Although *GEDICorrect* implements multiple waveform similarity metrics, this study focused on Kullback–Leibler (KL) divergence to demonstrate the framework's capabilities relative to baseline GEDI data and the standard *GEDI Simulator* correction method. A systematic evaluation of all available metrics within *GEDICorrect*, applied consistently across orbit-, beam-, and footprint-level corrections and extended to large study areas with diverse land-cover types and terrain conditions, represents a substantial effort deserving a dedicated analysis. Such an investigation will be presented separately, ensuring that the present work remains focused on describing the tool and demonstrating its effectiveness for GEDI geolocation correction. By enhancing geolocation precision in a scalable and computationally efficient way, *GEDICorrect* opens new opportunities for more accurate assessments of canopy structure and terrain. This capability has broad relevance, from improving biomass estimation and carbon accounting to supporting biodiversity monitoring and conservation planning at regional to global scales.

**CRediT authorship contribution statement**

**Leonel Corado**: Conceptualization, Methodology, Software, Validation, Formal analysis, Investigation, Data curation, Visualization, Writing – original draft, Writing – review & editing.

**Sérgio Godinho**: Conceptualization, Methodology, Validation, Formal analysis, Investigation, Data curation, Visualization, Writing – original draft, Writing – review & editing, Funding acquisition, Supervision, Project administration.

**Carlos A. Silva**: Conceptualization, Writing – review & editing

**Juan Guerra-Hernández**: Conceptualization, Writing – review & editing

**Francesco Valério**: Visualization, Writing – review & editing

**Teresa Gonçalves:** Resources, Writing – review & editing

**Pedro Salgueiro:** Resources, Writing – review & editing

**Data and Code Availability**

The framework developed and used for the experiments, along with detailed installation and execution instructions, is available in the GitHub repository: https://www.github.com/leonelluiscorado/GEDICorrect/. An example dataset (including GEDI and ALS data) is provided in the repository to illustrate the workflow. The complete dataset used in this study comprises ~200 GB of GEDI and ALS (.laz) files; due to its large size, these data are not hosted online but can be made available upon request.

**Declaration of Generative AI in Scientific Writing**

During the preparation of this work the authors used OpenAI's ChatGPT in order to improve readability and correct English grammar and spelling. After using this tool/service, the authors reviewed and edited the content as needed and take full responsibility for the content of the published article.

**Acknowledgments**

This study, including the research contract awarded to Leonel Corado, was funded by the GEDI4SMOS project (*Combining LiDAR, radar, and multispectral data to characterize the three-dimensional structure of vegetation and produce land cover maps*), financially supported by the Directorate-General for Territory (DGT) through the Recovery and Resilience Plan (*Investimento RE-C08-i02: Cadastro da Propriedade Rústica e Sistema de Monitorização da Ocupação do Solo*). The authors also acknowledge the R&D unit MED – Mediterranean Institute for Agriculture, Environment and Development (https://doi.org/10.54499/UIDB/05183/2020; https://doi.org/10.54499/UIDP/05183/2020) and the Associate Laboratory CHANGE – Global Change and Sustainability Institute (https://doi.org/10.54499/LA/P/0121/2020) for institutional support. Furthermore, the authors also acknowledge the R&D unit Algoritmi (https://doi.org/10.54499/UIDB/00319/2020) and the associate laboratory LASI - Intelligent Systems Associate Laboratory (https://doi.org/10.54499/LA/P/0104/2020) for institutional support.